\documentclass[journal]{IEEEtran}
\usepackage{graphicx,times,cite,amsmath, amssymb, epsfig, array,comment}
\usepackage{multirow}
\usepackage[noend]{algorithmic}
\usepackage{algorithm}

\usepackage{array}
\usepackage{textcomp}
\newlength{\figwidth}
\begin{document}
\title{\LARGE A Hybrid Advertising Mode for Device Discovery \\in Bluetooth Low Energy Networks}
\author{Zhong~Shen, Hai~Jiang, Rongfei Fan, and Hongxing Guo\vspace{-7mm}\thanks{Z. Shen is with the School of Telecommunications Engineering, Xidian University, Xi’an 710071, China, and is also with Guangzhou Institute of Technology, Xidian University, Guangzhou, Guangdong 510555, China (e-mail: zhshen@mail.xidian.edu.cn). H. Jiang is with the Department of Electrical and Computer Engineering, University of Alberta, Edmonton, AB, Canada T6G 2V4 (e-mail: hai1@ualberta.ca). R. Fan is with the School of Cyberspace Science and Technology, Beijing Institute of Technology, Beijing 100081, P. R. China (e-mail: fanrongfei@bit.edu.cn). H. Guo is with the School of Computer Science and Technology, Huazhong University of Science and Technology, Wuhan 430074, China (e-mail: guohx@mail.hust.edu.cn).}
}

\maketitle
\begin{abstract}
Device discovery has a great impact on the performance of Bluetooth low energy (BLE). The performance of device discovery is highly related to the advertising mode. BLE has two advertising modes: pseudo-random delay advertising (RDA) and periodic deterministic advertising (PDA). Generally, PDA has low discovery latency but is susceptible to persistent collisions, whereas RDA does not suffer persistent collisions but has much larger discovery latency. In this paper, we propose a novel hybrid advertising mode, called Deterministic and pseudo-Random delay Advertising (DRA), which has the advantages of both PDA and RDA. We develop an analytical model for DRA mode based on a two-dimensional discrete-time Markov chain, and analyze the expected discovery latency of DRA in a single-advertiser case and a multiple-advertiser case. Simulation shows the accuracy of our analytical model, and also verifies that DRA can achieve an excellent tradeoff between low discovery latency and robustness to collisions.
\end{abstract}

\section{Introduction}
Due to extremely low energy consumption, Bluetooth Low Energy (BLE) is popular for short-range wireless communications and is available in almost all smart devices, such as smartphones, smart watches, and electronic products. 
Device discovery is a prerequisite for BLE communications and has a great impact on the performance of BLE. According to BLE specification, for a device to discover another device, one device must act as a \emph{scanner}, and the other must act as an \emph{advertiser}. BLE specifies three advertising channels, called channel 37, 38, and 39, over 2.4 GHz Industrial, Scientific and Medical (ISM) bands. An advertiser broadcasts advertisements over the three advertising channels. A scanner tries to discover advertisers by scanning the three advertising channels to receive advertisements of the advertisers.

\begin{figure*}
\centering
\includegraphics[scale = 0.7]{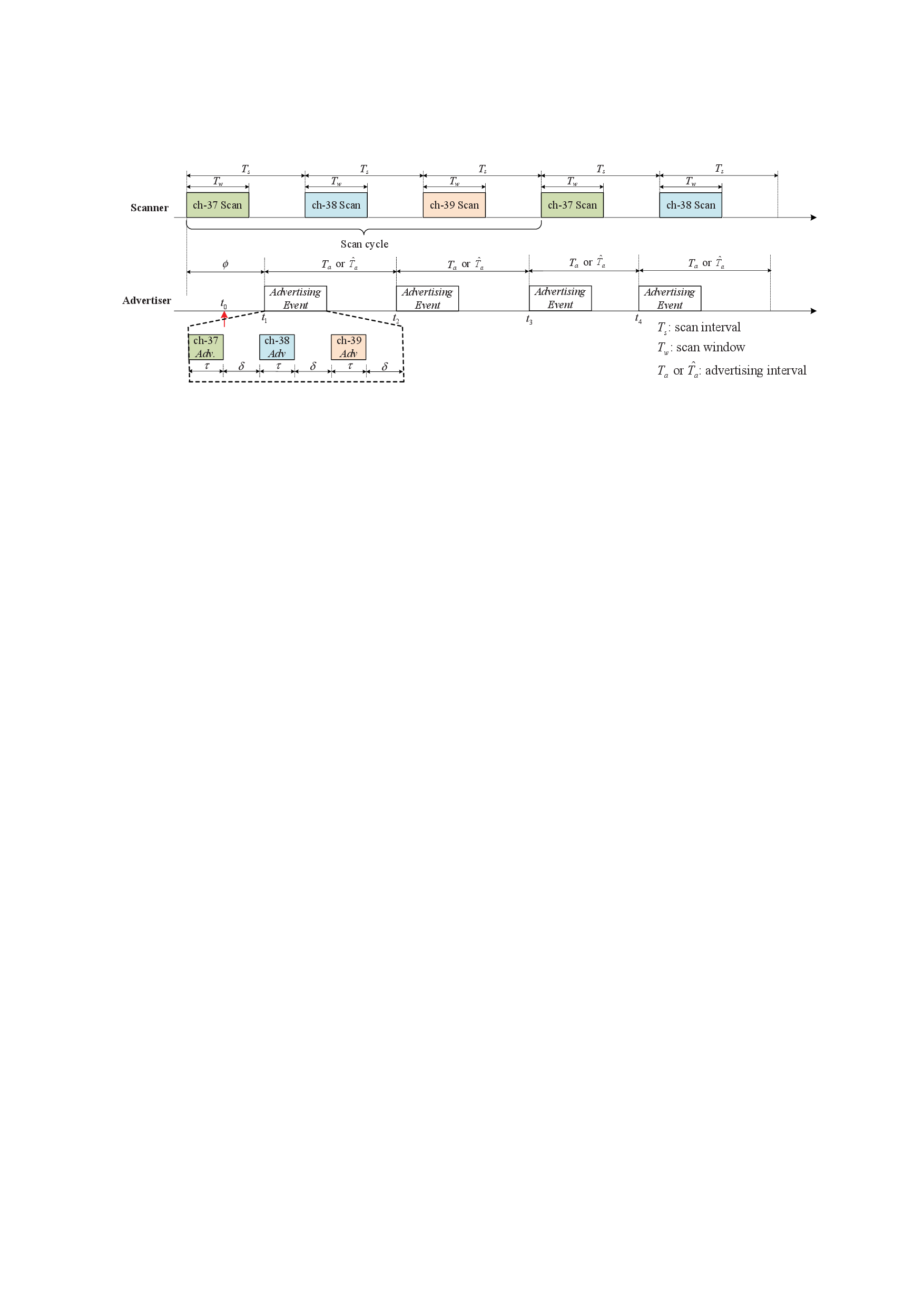}
\caption{BLE scanning and advertising.}\label{draexample}
\end{figure*}

The BLE device discovery process is specified in Generic Access Profile (GAP) of BLE. Fig.~\ref{draexample} shows an example of a scanner and an advertiser. An advertising event (advEvent) is periodically initiated by the advertiser. Each advEvent contains three advertising Packet Data Units (Adv\underline{\space}PDUs), sequentially sent over the three advertising channels. The duration for an Adv\underline{\space}PDU transmission is $\tau$. After transmitting an Adv\underline{\space}PDU on one advertising channel, the advertiser will wait for duration $\delta$ on the same channel for responses. As for the scanner, it scans channel 37, 38, and 39 in turn to receive Adv\underline{\space}PDUs from the advertiser. The scanner performs a scan every $T_s$ duration (called {\it scan interval}). In specific, there is a {\it scan window} with duration $T_w$ at the beginning of a scan interval. The scanner scans one of the advertising channels within the scan window, and keeps idle until the end of the scan interval. The procedure is repeated such that the three advertising channels are scanned in turn.

Define {\it advertising interval} as the interval from the beginning of an advEvent to the beginning of the next advEvent. BLE has two advertising modes: the pseudo-random delay advertising (RDA) (specified in Bluetooth 4.2 \cite{Ble4}), and the periodic deterministic advertising (PDA) (specified in Bluetooth 5.0 \cite{Ble5.0}). The advertising interval in RDA, denoted by $\hat{T}_a$, is the summation of a fixed duration $T_\ell$ and a pseudo-random duration $T_d$, while the advertising interval in PDA, denoted by $T_a$, is a fixed duration. In PDA mode and RDA mode, we define {\it advertising interval parameter (AIP)} of an advertiser as the duration and expected (mean) duration, respectively, of the advertising interval.


There are research efforts in the literature that analyze BLE device discovery \cite{LiuCommLett2012, ChoSensors2015, JulieIPSN2017, JeonTVT2017, LuoWirel2020, ShenTMC2021}. The works in \cite{LiuCommLett2012, ChoSensors2015, JulieIPSN2017} give analytical models for RDA, by assuming that the scan window of the scanner is at least the size of the advertising interval of the advertiser (i.e., $T_w \ge \hat{T}_a$). However, this assumption may not be practical. The works in \cite{JeonTVT2017, LuoWirel2020} use Chinese Remainder Theorem to analyze discovery latency, by approximating each duration parameter to an integer number of slots.  However, this approximation may reduce the accuracy of analytical models. The work in \cite{ShenTMC2021} develops an analytical model for PDA mode and an analytical model for RDA mode. 

It is shown in \cite{ShenTMC2021} that for advertisers with the same AIP, PDA has lower discovery latency than RDA. On the other hand, PDA is susceptible to persistent collisions, as follows. For two advertisers using PDA, if their advertising intervals are equal and their initial Adv\underline{\space}PDU transmissions over an advertising channel overlap, then their Adv\underline{\space}PDUs will collide repeatedly. RDA does not suffer persistent collisions, thanks to the pseudo-random duration in each advertising interval.

Motivated by the above advantages and disadvantages of PDA and RDA, we propose a novel hybrid advertising mode, called Deterministic and pseudo-Random delay Advertising (DRA), which combines the advantages of both PDA and RDA. The DRA works as follows. For every $m$ advertising intervals of an advertiser, the advertiser works in PDA mode for $m-1$ advertising intervals, and works in RDA mode for the remaining advertising interval. We build an analytical model for the DRA mode, and analyze the expected discovery latency of DRA in a single-advertiser case and a multiple-advertiser case. Simulation shows the accuracy of our analytical model, and verifies that DRA can achieve an excellent tradeoff between low discovery latency and robustness to collisions.


\section{Proposed DRA Mode}\label{s:DRA}
For PDA mode, the advertising interval, denoted by $T_a$, is deterministic. For RDA mode, however, the advertising interval, denoted by $\hat{T}_a$, is a random variable consisting of a fixed duration ${T_\ell }$ and a pseudo-random duration $T_d$. As per BLE specification, $T_d$ is uniformly distributed within the range 0 ms to 10 ms. Without loss of generality, we consider $T_d$ as an integer random variable that is  uniformly distributed within range $\{0,1,2,...,r\}$. 

DRA is a hybrid mode combining PDA mode with advertising interval $T_a$ and RDA mode with advertising interval $\hat{T}_a$, in which the mean of $\hat{T}_a$ is equal to $T_a$. Define $T_a $ and the mean of $\hat{T}_a$ as AIP of a DRA-mode advertiser. The main idea of DRA mode is as follows. Among every $m$ advertising intervals (with $m$ being a design parameter for DRA), the advertiser works in PDA for the first $m-1$ advertising intervals, and works in RDA for the last advertising interval.

For an advertiser to implement DRA, each advertising interval is associated with a mode index to indicate which advertising mode (PDA or RDA) will be used. Suppose the mode index of current advertising interval is $i$ ($m - 1 \ge i \ge 0$). The mode index of the next advertising interval will be $(i + 1)\bmod m$. For an advertising interval with a nonzero mode index, PDA is used (i.e., the advertising interval is $T_a$). For an advertising interval with mode index being zero, RDA is used (i.e., the advertising interval is $\hat{T}_a$).


\section{Expected Discovery Latency of DRA Mode with a Single Advertiser}\label{s:rda}
\begin{figure*}
\centering
\includegraphics[scale = 0.3]{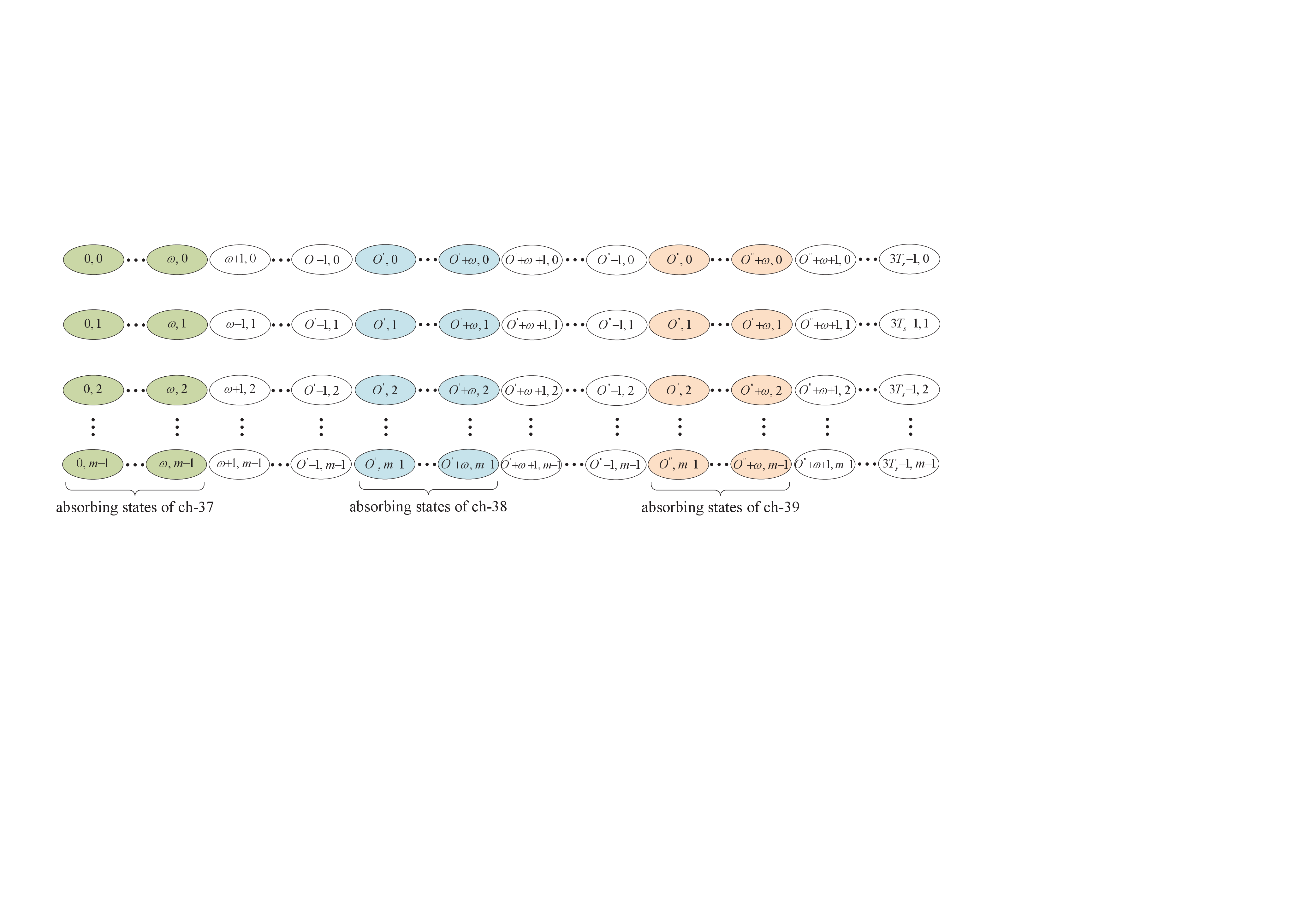}
\caption{States of the two-dimensional discrete-time Markov chain for DRA mode.}\label{2DMarkov}
\end{figure*}

We first build an analytical model for DRA mode with a single advertiser. Suppose a scanner is within the transmission range of the advertiser starting from moment $t_0$, as shown in Fig.~\ref{draexample}. After time instant $t_0$, the mode index of the first complete advertising interval is assumed to be 1. So the subsequent advertising intervals have mode index $2,3,...,m-1,0,1,2...$. Denote $t_j$ as the moment at which the \emph{j}th advertising interval begins. $t_j$ is also the time instant at which the channel-37 Adv\underline{\space}PDU of the \emph{j}th advEvent is sent. The \emph{time offset} of the \emph{j}th advertising interval is defined as the time difference between $t_j$ and the time instant at which the scanner starts to scan channel 37 right before $t_j$. The initial time offset is the time offset of the first advertising interval, denoted by $\phi$, as shown in Fig.~\ref{draexample}.

Next, we investigate the discovery procedure from the perspective of time offsets of the advertising intervals. The scan interval of the scanner is $T_s$ time units, as shown in Fig.~\ref{draexample}. Recall that the three advertising channels are scanned in turn. Thus, the scanner's scan cycle has $3T_s$ time units, as shown in Fig.~\ref{draexample}, and the time offsets of the advertiser's advertising intervals range from 0 to $3T_s-1$ time units. Initially, the time offset of the first advertising interval is $\phi$. The time offset of the second advertising interval is $(\phi  + {T_a})\bmod 3{T_s}$ if the mode index of the first advertising interval is not zero, or is $(\phi  + \hat{T}_a)\bmod 3{T_s}$ otherwise. The time offsets of all subsequent advertising intervals will change according to the same rule. It can be seen from Fig.~\ref{draexample} that if the time offset of an advertising interval is within 0 to $T_w-\tau$, then the transmission of the advertiser's Adv\underline{\space}PDU on channel 37 of the current advertising interval will be completely within the scanner's scan window on channel 37, i.e., the scanner discovers the advertiser on channel 37. Similarly, if the time offset of the current advertising interval is from ${T_s} - \tau  - \delta$ to $({T_s} - \tau  - \delta)+T_w-\tau$ (or from $2({T_s} - \tau  - \delta)$ to $2({T_s} - \tau  - \delta)+T_w-\tau$), then the transmission of the advertiser's Adv\underline{\space}PDU on channel 38 (or channel 39) of the current advertising interval will be entirely within the scanner's scan window on channel 38 (or channel 39), i.e., the scanner discovers the advertiser on channel 38 (or channel 39). 
The discovery latency is defined as the delay from the start of the advertiser's first advertising interval to the time when a discovery occurs.

To compute the discovery latency, we build a two-dimensional discrete-time Markov chain. Denote a state of the Markov chain as $(x,y)$, $x \in \{0,1, \cdots ,3{T_s} - 1\}$, $y \in \{0,1, \cdots ,m - 1\}$. Here $x$ represents the time offset of the advertiser's advertising interval, and $y$ represents the mode index of the advertiser. Define {\it absorbing state} as a state that discovery happens. State $(x,y)$ is an absorbing state if $x\in \{0,1, \cdots ,\omega\} $ ($\omega  \triangleq {T_w} - \tau $) because at the state, the scanner will discover the advertiser on channel 37. Similarly, State $(x,y)$ is also an absorbing state if $x \in \{ {O'},O'+1, \cdots ,{O'} + \omega \}$ with ${O'} \triangleq {T_s} - \tau  - \delta$ (or $x \in \{ {O''},O''+1, \cdots ,{O''} + \omega \}$ with ${O''} \triangleq 2({T_s} - \tau  - \delta )$) because at the state, the scanner will discover the advertiser on channel 38 (or channel 39). Fig.~\ref{2DMarkov} shows states of the Markov chain, including absorbing states for channel 37, 38, and 39, and other states (called {\it transient states}).

\begin{figure}
\centering
\includegraphics[scale = 0.3]{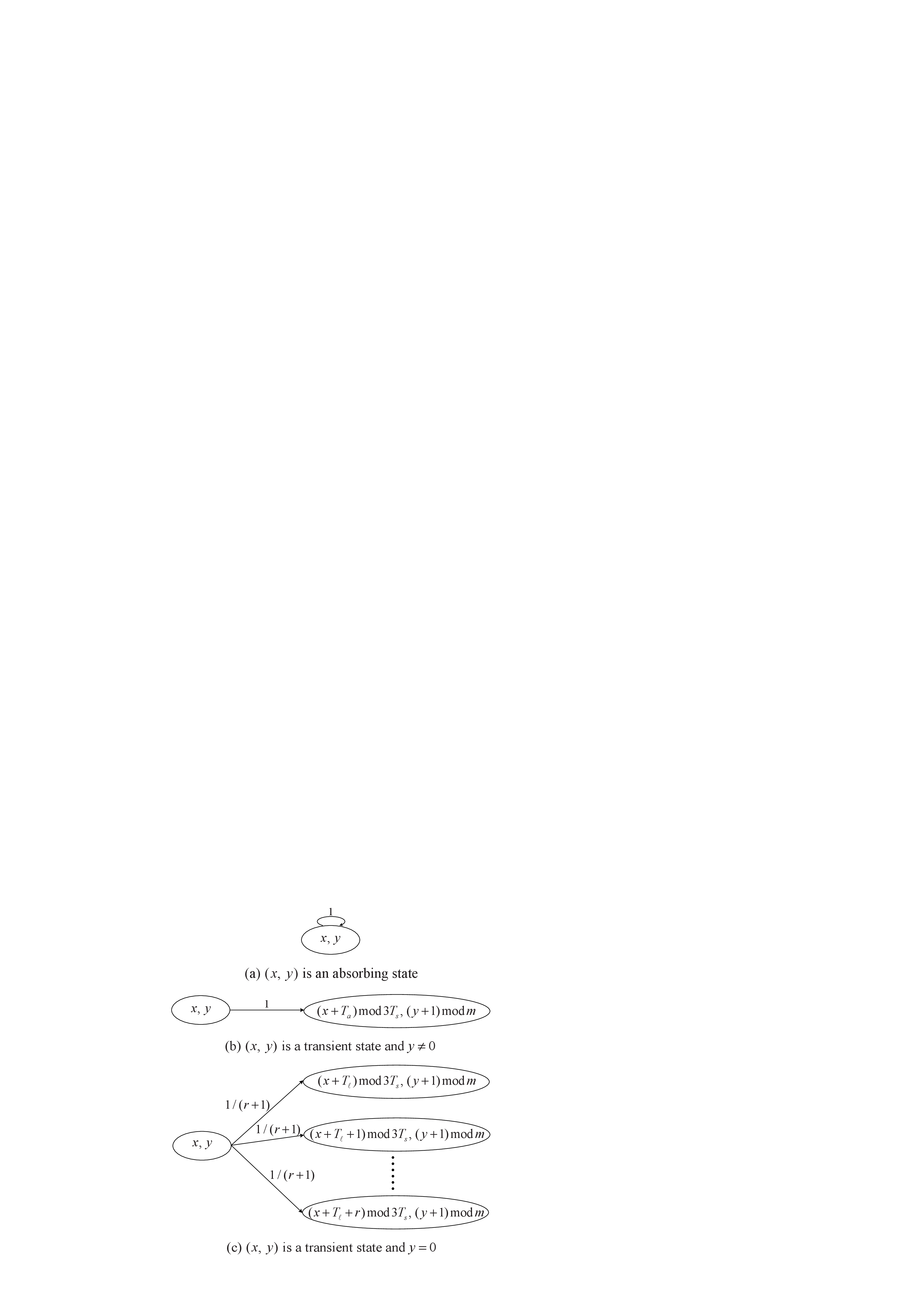}
\caption{Transition probability graph of the two-dimensional discrete-time Markov chain.}\label{transition}
\end{figure}

We sample the state of the system over each advertising interval of the advertiser. The state evolution forms a two-dimensional discrete-time Markov chain.
Fig.~\ref{transition} shows the transition probability graph of the two-dimensional discrete-time Markov chain. For an absorbing state $(x,y)$, its transition probability is $1$ to itself, and is $0$ to any other state, as shown in Fig.~\ref{transition}(a). For a transient state $(x,y)$:
\begin{itemize}
\item If $y \ne 0$ (i.e., the advertiser works in PDA mode), the time offset of the next advertising interval is $(x + {T_a})\bmod 3{T_s}$, and the mode index of the next advertising interval is $(y + 1)\bmod m$. In other words, the next state is $\big( (x + {T_a})\bmod 3{T_s},  (y + 1)\bmod m \big)$, as shown in  Fig.~\ref{transition}(b).

 \item If $y = 0$, it means that the advertiser is in RDA mode. Recall that the pseudo-random component ($T_d$) of the advertising interval $\hat{T}_a$  is an integer  random variable that is  uniformly distributed within range $\{0,1,2,...,r\}$. So the time offset of the next advertising interval can be $(x + {T_\ell })\bmod 3{T_s}$, $(x + {T_\ell } + 1)\bmod 3{T_s}$, $ \cdots $, or $(x + {T_\ell } + r)\bmod 3{T_s}$, each with probability of $\frac{1}{{r + 1}}$, and the mode index of the next advertising interval will be $(y + 1)\bmod m$, as shown in Fig.~\ref{transition}(c).
\end{itemize}
Let ${p_{(x,y),({x'},{y'})}}$ denote transition probability from state $(x,y)$ to $(x',y')$, and ${\mu _{(x,y)}}$ denote the expected number of transitions for the Markov chain to evolve from state $(x,y)$ to an absorbing state. We have
\begin{align}\label{equ:avg_discovery_lat}
\ &\begin{array}{l}
{\mu _{(x,y)}}=0, {{\rm{~~~~~~~~~~~if~}} (x,y) {\rm{~is~an~absorbing~state}}},\\
{\mu _{(x,y)}}=1 + \sum\limits_{{x'} = 0}^{3{T_s} - 1}{\sum\limits_{{y'} = 0}^{m - 1} {{p_{(x,y),({x'},{y'})}}} }{\mu _{({x'},{y'})}},\\
{{\rm{~~~~~~~~~~~~~~~~~~~~~~~~if~}} (x,y) {\rm{~is~a~transient~state}}}.
\end{array}
\end{align}
Based on (\ref{equ:avg_discovery_lat}), we can calculate ${\mu _{(x,y)}}$ for each state $(x,y)$. Then the expected discovery latency of the scanner, denoted as $X$, is given as
\begin{equation}\label{e:DRA_exp_dis_latency}
X = \frac{T_a}{{3{T_s} \times m}}\sum\limits_{x = 0}^{3{T_s} - 1} {\sum\limits_{y = 0}^{m - 1} {{\mu _{(x,y)}}}}.
\end{equation}

\section{Expected Discovery Latency of DRA Mode with Multiple Advertisers}\label{s:explatency}
In this section, we analyze the discovery performance of DRA with multiple advertisers. Consider a target scanner, a target advertiser, and a number of surrounding advertisers. All the advertisers adopt DRA mode. Recall that for a DRA-mode advertiser, the $T_a$ in PDA mode and the mean of $\hat{T}_a$ in RDA mode are equal and are called the AIP of the advertiser. 

Consider the two-dimensional Markov chain for the target scanner to discover the target advertiser. Suppose an absorbing state is achieved, i.e., on an advertising channel, an Adv\underline{\space}PDU of the target advertiser (called the target Adv\underline{\space}PDU) is completely included within a scan window of the target scanner. If one surrounding advertiser (called colliding advertiser) also transmits Adv\underline{\space}PDU on that advertising channel and the Adv\underline{\space}PDU overlaps with the target Adv\underline{\space}PDU, a collision occurs, and we call the absorbing state a {\it collided absorbing state}. So the Markov chain will continue to evolve from the collided absorbing state to next states. If the target and colliding advertisers have different AIPs, then the collision is called a {\it non-persistent collision}, because it is likely that when the system enters the next absorbing state, the colliding advertiser will not collide with the target advertiser again. However, if the target and colliding advertisers have the same AIP and both work in PDA mode, then the collision will continue for a period of time. The collision may be resolved when either the target or the colliding advertiser enters RDA mode, thanks to the pseudo-random delay component of the advertising interval in RDA mode. We call this kind of collision \emph{temporary persistent collision}. Next, we analytically derive the expected discovery latency with possible collisions.

Denote the AIP of the target advertiser as $T_a$. Among all surrounding advertisers, totally $N_1$ advertisers have their AIPs different from $T_a$, called group-1 advertisers, and totally $N_2$ advertisers have their AIPs being equal to $T_a$, called group-2 advertisers. For the $N_1$ group-1 advertisers, denote their AIPs as $T_{a,1}, T_{a,2},...,T_{a,N_1}$. So if a group-1 advertiser collides with the target advertiser, it is a non-persistent collision. If a group-2 advertiser collides with the target advertiser, it is a temporary persistent collision.

Consider an absorbing state of the two-dimensional Markov chain for the target scanner to discover the target advertiser. At the absorbing state, the probability that the $i$th group-1 advertiser collides with the target advertiser is $\frac{2\tau}{T_{a,i}}$, and the probability that a group-2 advertiser collides with the target advertiser is $\frac{2\tau}{T_{a}}$. So overall, the probability of no collision at the absorbing state is
\begin{equation}\label{e:nc_prob}
P_\text{no-col}=\left[\prod\limits_{i = 1}^{N_1} {(1 - \frac{{2\tau }}{{T}_{a,i}})}\right] (1-\frac{2\tau}{T_a})^{N_2}.
\end{equation}

If a collision happens at the absorbing state, the absorbing state is a collided absorbing state. Next we discuss what happens if the system enters a collided absorbing state, say state $(x,y)$. So, the system should treat $(x,y)$ as a transient state and push it to the next state, say $(x',y')$, according to the transition probability graph shown in Fig.~\ref{transition}(b) and \ref{transition}(c). Denote $K_{(x,y)}$ as the number of transitions from collided absorbing state $(x,y)$ to the next absorbing state.

If $y\neq 0$, the next state $(x',y')$ only has one option as shown in Fig.~\ref{transition}(b), and we have $K_{(x,y)} = 1 + \mu_{(x',y')}$, with $x' = x + T_a \bmod 3T_s, ~y'= (y+1) \bmod m$. If $y = 0$, the next state $(x',y')$ has $r+1$ options as shown in Fig.~\ref{transition}(c), as $(x'_{0},y'), (x'_1,y'),...,(x'_r,y')$ with $x'_j = x + T_\ell + j \bmod 3T_s,~ y'=y+1 \bmod m$, with $j\in \{0,1,2,...,r\}$. So we have
\[K_{(x,y)} = \frac{1}{r+1}\sum_{j=0}^r (1 + \mu_{(x'_j, y')}).\]
Thus, the average number of transitions from a collided absorbing state to the next absorbing state is given as
\begin{equation}\label{e:gamma_ex}
\gamma = \frac{1}{3(\omega +1)m}\sum_{\text{$(x,y)$ is absorbing state}} K_{(x,y)}.
\end{equation}
Here the term $\frac{1}{3(\omega +1)m}$ is due to the fact that the collided absorbing state $(x,y)$ could be any one of the totally $3(\omega+1)m$ absorbing states of the Markov chain.

Denote $H$ as the expected duration from a collided absorbing state to discovery (i.e., to an absorbing state with no collision). Then, the expected discovery latency for the target advertiser is given as
\begin{equation}\label{e:exp_latency_col}
X + (1- P_\text{no-col}) H
\end{equation}
with $X$ given in (\ref{e:DRA_exp_dis_latency}). Next we derive an expression of $H$.

Denote $H_1$ and $H_2$ as the expected duration from a collided absorbing state that is caused by a collision with a group-1 advertiser and a group-2 advertiser, respectively, to the discovery of the target advertiser. We have
\begin{equation}\label{e:H}
H = P_1 H_1 + (1-P_1) H_2
\end{equation}
with $P_1 = \frac{\sum_{i=1}^{N_1} \frac{2\tau}{T_{a,i}}  }{\sum_{i=1}^{N_1} \frac{2\tau}{T_{a,i}} + N_2 \frac{2\tau}{T_a}}$ being the probability that a collided absorbing state is caused by a collision with a group-1 advertiser.

From a collided absorbing state caused by a collision with a group-1 advertiser, it takes (on average) $\gamma$ transitions to reach the next absorbing state ($\gamma$ given in (\ref{e:gamma_ex})). If the next absorbing state does not have collision (with probability $P_\text{no-col}$), then discovery happens; otherwise, we need duration $H$ to reach discovery. Thus, we have
\begin{equation}\label{e:H1}
H_1 = \gamma T_a + (1 - P_\text{no-col}) H.
\end{equation}

If a collided absorbing state is caused by a collision with a group-2 advertiser, this collision is a temporary persistent collision.
Suppose the temporary persistent collision happens when the mode indexes of the target advertiser and the colliding advertiser is $u$ and $v$, respectively, $u \in \{0,1, \cdots ,m - 1\}$, $v \in \{0,1, \cdots ,m - 1\}$. So, it takes the target advertiser $(m - u)$ transitions such that the target advertiser enters RDA mode. Similarly, it takes the colliding advertiser $(m - v)$ transitions such that the colliding advertiser enters RDA mode. Since the added random delay (with mean value of 5 ms) in RDA mode is much larger than $2\tau$ (no more than 800 $\mu$s \cite{ShenTMC2021}), once the target advertiser or the colliding advertiser enters RDA mode, it is considered that the temporary persistent collision has been resolved. Before either advertiser enters RDA mode, any absorbing state for the target advertiser will be a collided absorbing state. Thus, the Markov chain for the target advertiser will re-enter collided absorbing state(s) $\left\lfloor {\frac{{\min \{m - u,m - v\}}}{{{\gamma}}}} \right\rfloor $ times ($\lfloor \cdot \rfloor$ being the floor function). Denote ${R}$ as the expected duration for re-entering those collided absorbing states, we have
\begin{equation}\label{e:tpc_duration}
R = \frac{{{\gamma } \times {T_a}}}{{{m^2}}}\sum\limits_{u = 0}^{m - 1} {\sum\limits_{v = 0}^{m - 1} {\left\lfloor {\frac{{\min \{m - u,m - v\}}}{{{\gamma }}}} \right\rfloor } }.
\end{equation}

Thus, we have
\begin{equation}\label{e:H2}
H_2 = R + \gamma T_a  + (1 - P_\text{no-col}) H.
\end{equation}

From (\ref{e:H}), (\ref{e:H1}), and (\ref{e:H2}), we have
\begin{equation}\label{e:H_exp}
H = \frac{\gamma T_a + (1-P_1)R}{P_\text{no-col}}.
\end{equation}
Thus,  the expected discovery latency for the target advertiser is given as in (\ref{e:exp_latency_col}), with $H$ given as (\ref{e:H_exp}).

%

\section{Performance Evaluation}\label{s:eva}
We use a customized C++ simulator to simulate expected discovery latency of our proposed DRA mode and compare with our analytical expressions of the expected discovery latency. In simulation, advertisers use a data rate of 1 Mbps to broadcast non-connectable undirected advertising PDUs, i.e., ADV\underline{\space}NONCONN\underline{\space}IND. We set $\tau = 376$ $\mu$s and $\delta = 437$ $\mu$s, as indicated in \cite{ShenTMC2021}. For RDA mode, the pseudo-random delay $T_d$ is an integer random variable that is uniformly distributed over the interval from 0 to $10,000\ \mu$s. As suggested by \cite{ShenTMC2021}, we set $T_s=310,000 \ \mu$s and $T_w = 10,375\ \mu $s in all simulation, unless otherwise specified. Each simulated value is the average of 1,000,000 simulation runs.



\begin{figure}
\centering
\includegraphics[scale = 0.20]{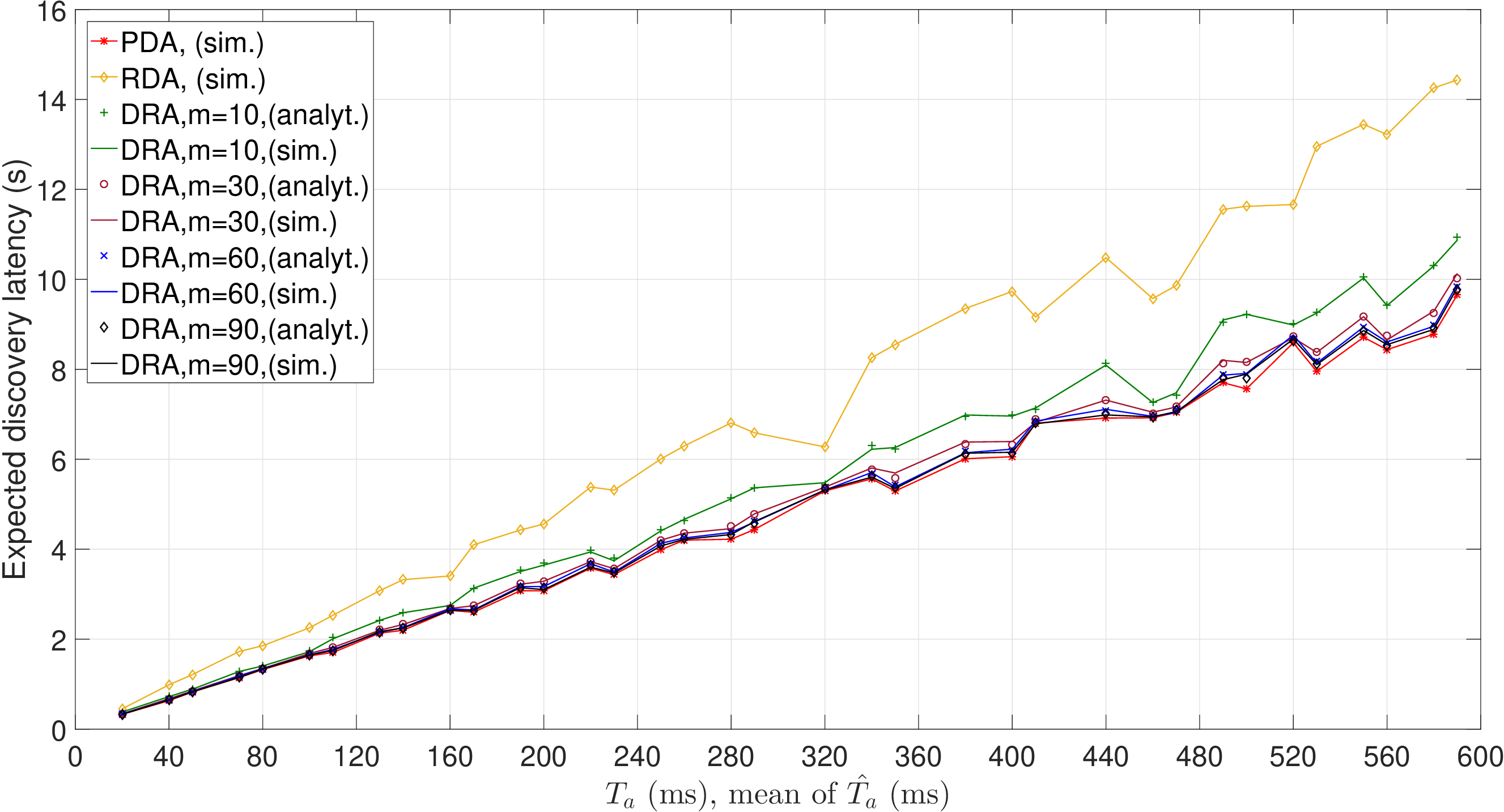}
\caption{Expected discovery latency in DRA mode, PDA mode, and RDA mode with a single advertiser.}\label{dra_latency}
\end{figure}

We first study the performance of DRA mode with a single advertiser. Fig.~\ref{dra_latency} shows the analyzed and simulated expected discovery latency of DRA mode when the advertiser's AIP ($T_a$ and the mean of $\hat{T}_a$) changes from 20,000 $\mu $s to 590,000 $\mu$s. It is seen that with different values of $m$, our analytical results that are calculated by  (\ref{e:DRA_exp_dis_latency}) and the simulated values match well, which verifies the correctness of our analysis model. For comparison purpose, the simulated expected discovery latency of PDA mode with different AIP ($T_a$) and RDA mode with different AIP (the mean of $\hat{T}_a$) are also shown in Fig.~\ref{dra_latency}. As expected, DRA's expected discovery latency is in between those of the PDA and RDA. DRA's performance is closer to that of RDA when $m$ is small. As $m$ increases, the performance of DRA is approaching to that of PDA.

\begin{figure}
\centering
\includegraphics[scale = 0.20]{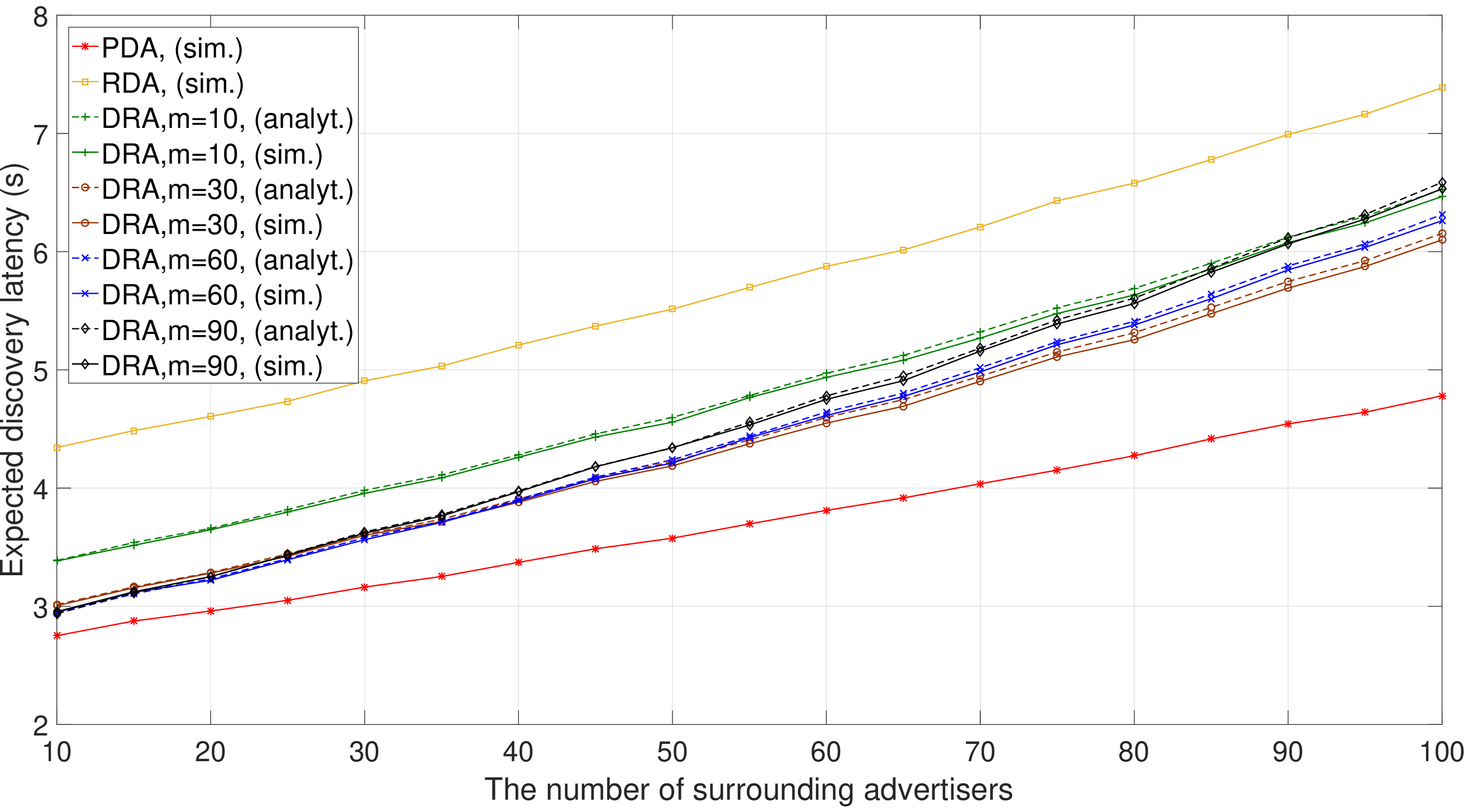}
\caption{Expected discovery latency in DRA mode, PDA mode (when persistent collision does not happen), and RDA mode with multiple advertisers.}\label{dra_collision_latency}
\end{figure}

Next, we study the performance of DRA mode with interference by surrounding advertisers. The AIP of the target advertiser is set to $170 {\rm ms}$. For each surrounding advertiser, its AIP is uniformly selected from $\{100 {\rm ms}, 170 {\rm ms}, 250 {\rm ms}\}$. Fig.~\ref{dra_collision_latency} shows the analyzed and simulated expected discovery latency when all advertisers adopt DRA mode and the number of surrounding advertisers ranges from 10 to 100. It is seen that the analytical and simulation results match well. For comparison, we also simulate the cases when all advertisers adopt PDA and when all advertisers adopt RDA. The simulated expected discovery latency values are also shown in Fig.~\ref{dra_collision_latency}. Note that when all advertisers adopt PDA, if the target advertiser collides with a surrounding advertiser whose AIP is also $170 {\rm ms}$, then the collision will be persistent (i.e., the two advertisers will collide forever), and the discovery latency will become infinite. In particular, when the number of surrounding advertisers changes from 10 to 100 (with step size 5), the probability of persistent collision in our simulation is 1.19\%, 2.07\%, 2.49\%, 3.38\%, 4.24\%, 4.66\%, 5.49\%, 6.30\%, 6.78\%, 7.61\%, 8.34\%, 8.78\%, 9.43\%, 10.32\%, 10.68\%, 11.39\%, 12.31\%, 12.64\%, and 13.42\%, respectively.
Thus, in Fig.~\ref{dra_collision_latency}, for the PDA curve (i.e., all advertisers adopt PDA), each simulated value is the average of the simulation runs in which a persistent collision does not happen.
It can be seen that, when persistent collision does not happen, PDA has the lowest discovery latency. RDA does not suffer persistent collision, but has the largest discovery latency.  Our proposed DRA achieves a better tradeoff between PDA and RDA. In other words, DRA does not suffer persistent collision, and has lower expected discovery latency than RDA. 


\section{Conclusions}\label{s:con}
In this paper, we propose the DRA, to take the advantages of both PDA and RDA. We develop an analytical model for DRA mode based on a two-dimensional discrete-time Markov chain, and analyze the expected discovery latency of DRA for a single-advertiser case and for a multiple-advertiser case. It is shown that DRA can achieve excellent tradeoff between low discovery latency and robustness to collisions.


\clearpage

\end{document}